\documentclass[prd,twocolumn,superscriptaddress,showpacs]{revtex4-1}
\usepackage{graphicx,amsmath,amssymb,bm}

\newcommand{\kf}{k_{\rm F}}
\newcommand{\fmiq}{\, \text{fm}^{-3}}
\newcommand{\mev}{\, \text{MeV}}

\begin{document}

\title{Spin-dependent WIMP scattering off nuclei}

\author{J.\ Men\'{e}ndez}
\affiliation{Institut f\"ur Kernphysik, 
Technische Universit\"at Darmstadt, 
64289 Darmstadt, Germany}
\affiliation{ExtreMe Matter Institute EMMI, 
GSI Helmholtzzentrum f\"ur Schwerionenforschung GmbH, 
64291 Darmstadt, Germany}
\author{D.\ Gazit}
\affiliation{Racah Institute of Physics and
The Hebrew University Center for Nanoscience and Nanotechnology, 
The Hebrew University, 91904 Jerusalem, Israel}
\author{A.\ Schwenk}
\affiliation{ExtreMe Matter Institute EMMI, 
GSI Helmholtzzentrum f\"ur Schwerionenforschung GmbH, 
64291 Darmstadt, Germany}
\affiliation{Institut f\"ur Kernphysik, 
Technische Universit\"at Darmstadt, 
64289 Darmstadt, Germany}

\begin{abstract}
Chiral effective field theory (EFT) provides a systematic expansion
for the coupling of WIMPs to nucleons at the momentum transfers
relevant to direct cold dark matter detection. We derive the currents
for spin-dependent WIMP scattering off nuclei at the one-body level
and include the leading long-range two-body currents, which are
predicted in chiral EFT. As an application, we calculate the
structure factor for spin-dependent WIMP scattering off
$^{129,131}$Xe nuclei, using nuclear interactions that have been
developed to study nuclear structure and double-beta decays in this
region. We provide theoretical error bands due to the nuclear
uncertainties of WIMP currents in nuclei.
\end{abstract}


\maketitle

\section{Introduction}

Cosmological and astrophysical observations have
established that more than $20 \%$ of the energy density of our
universe is dark matter, a rarely interacting nonbaryonic form of
matter, whose specific composition remains unknown~\cite{theo}.
Promising candidates are weakly interacting massive particles (WIMPs),
such as neutralinos, the lightest supersymmetric particles predicted
by extensions of the standard model. This has spurred direct detection
of cold dark matter via elastic scattering off nuclei, requiring detailed
knowledge of the response to WIMP induced currents in nuclei. This
presents a challenging problem, because even if the coupling of
neutralinos (or other particles) to quarks is known, it needs to be
evaluated at the nucleus level in the nonperturbative regime of
quantum chromodynamics~\cite{new}.

Chiral effective field theory (EFT)
provides a systematic expansion for nuclear forces and the
coupling to external probes for momenta of order of the pion mass, $p
\sim m_{\pi} \sim 100 \mev$, which are typical momentum transfers in
direct dark matter detection. In this paper, we focus on
spin-dependent neutralino scattering off nuclei, which is used to
constrain WIMP properties~\cite{exp} and is particularly sensitive to
nuclear structure. We derive the currents for spin-dependent WIMP
scattering at the one-body level and include the leading
long-range two-nucleon currents, which are predicted in chiral EFT.
Similar weak neutral currents are key for providing accurate
predictions of neutrino breakup of the deuteron for SNO~\cite{SNO},
while the corresponding weak charged currents have been found relevant
in Gamow-Teller and double-beta decays of medium-mass
nuclei~\cite{Javier}.

We apply the developed chiral EFT currents to calculate the structure
factor for spin-dependent WIMP scattering off $^{129,131}$Xe, as xenon
isotopes provide the tightest limits on WIMP couplings~\cite{XENON}.
To describe these nuclei, we employ the largest many-body spaces
accessible with nuclear interactions previously used to study nuclear
structure and double-beta decays in this region~\cite{Menendez}.

\section{Spin-dependent interactions}

We focus on spin-dependent WIMP
scattering, for which the low-momentum-transfer Lagrangian density
${\mathcal L}$ is taken to be a spatial axial-vector--axial-vector
coupling~\cite{IJMPE,Kam},
\begin{equation}
{\mathcal L} = - \frac{G_F}{\sqrt{2}} \:
\overline\chi {\bm \gamma} \gamma_5 \chi \cdot
\sum_q A_q \overline\psi_q {\bm \gamma} \gamma_5 \psi_q \,,
\label{L}
\end{equation}
where $G_F$ is the Fermi coupling constant, $\chi$ the neutralino
field, $\psi_q$ the fields of $q=u,d,s$ quarks, and $A_q$ are
neutralino-quark coupling constants. At the one-nucleon
level, the quark currents are replaced by their expectation value in a
single nucleon. This leads to one-body (1b) axial-vector currents
${\bf J}_{i,{\rm 1b}}$. Summing over all $A$ nucleons in a nucleus,
one has~\cite{IJMPE}
\begin{multline}
\sum_q A_q \overline\psi_q {\bm \gamma} \gamma_5 \psi_q
\longrightarrow \sum_{i=1}^A {\bf J}_{i,{\rm 1b}}
= \sum_{i=1}^A ( {\bf J}^0_{i,{\rm 1b}} + {\bf J}^3_{i,{\rm 1b}} ) \\
= \sum_{i=1}^A \frac{1}{2} \biggl[
a_0 \, {\bm \sigma}_i + a_1 \tau_i^3 \Bigl( {\bm \sigma}_i
- \frac{g_{P}(p^{2})}{2 m g_A} \,
({\bf p} \cdot {\bm \sigma}_{i}) \, {\bf p} \Bigl) \biggr] \,,
\end{multline}
where ${\bf p}={\bf p}_i-{\bf p}_i'$ denotes the momentum transfer
from nucleons (with mass $m$) to neutralinos. The isoscalar part ${\bf
J}^0_{i,{\rm 1b}}$ with coupling $a_0= (A_u+A_d)(\Delta u + \Delta
d) + 2A_s \Delta s$ receives contributions from the isoscalar
combination of the $u$ and $d$ quarks to the spin of the nucleon, as
well as from the $s$ quark. Here $\Delta q$ denotes the matrix element
of $\overline\psi_q {\bm \gamma} \gamma_5 \psi_q$ in the nucleon up to
the spin ${\bm \sigma}/2$~\cite{IJMPE}.
The isovector coupling can be written as
$a_1=(A_u-A_d) (\Delta u - \Delta d) = (A_u-A_d) g_A$, with the axial
coupling constant $g_A$. Therefore the isovector part ${\bf
J}^3_{i,{\rm 1b}}$ of the axial-vector WIMP coupling to the nucleon
is up to a factor $g_A$ identical to the axial-vector part of the weak
neutral current.

\section{Chiral EFT and WIMP currents}

The weak neutral current was
derived within chiral EFT for calculations of low-energy electroweak
reactions. At order $Q^0$ and $Q^2$, there are only
one-body currents, which for the isovector part of the axial-vector
WIMP current lead to
\begin{align}
{\bf J}^3_{i,{\rm 1b}} = \frac{1}{2} \,
a_1 \tau_i^3 \biggl( \frac{g_A(p^2)}{g_A} \, {\bm \sigma}_i
- \frac{g_{P}(p^{2})}{2 m g_A} \,
({\bf p} \cdot {\bm \sigma}_{i}) \, {\bf p} \biggl) \,.
\label{J3}
\end{align}
In chiral EFT, the $p$ dependence of the axial and
pseudo-scalar couplings $g_{A}(p^{2})$ and $g_{P}(p^{2})$ is due to
loop corrections and pion propagators, to order~$Q^2$~\cite{Bernard}:
\begin{equation}
\frac{g_{A}(p^{2})}{g_{A}} = 1-2 \, \frac{p^2}{\Lambda_{A}^2}
\hspace*{2.5mm} {\rm and} \hspace*{2.5mm}
g_P(p^2) = \frac{2 g_{\pi p n} F_\pi}{m_\pi^2+p^2} 
- 4 \, \frac{m g_A}{\Lambda_A^2} \,,
\label{gP}
\end{equation}
with $\Lambda_A = 1040 \mev$, pion mass $m_{\pi}=138.04 \mev$, pion
decay constant $F_\pi = 92.4 \mev$, and $g_{\pi p n} = 13.05$. In
previous calculations of WIMP scattering off nuclei, the smaller
$1/\Lambda_A^2$ terms are generally neglected and the
Goldberger-Treiman relation is implicitly used to write
$\frac{g_P(p^2)}{2 m g_A} \approx \frac{1}{m_\pi^2+p^2}$ (compare with
Ref.~\cite{IJMPE}).

Neglecting the strange quark contribution, the neutral axial-vector
currents are isovector in the Standard Model. Therefore, knowledge of
the isoscalar currents in the nucleon is based on models. These
suggest that the $Q^2$ one-body currents have a form-factor mass-scale
$\sim \Lambda_A$~\cite{isoscalar}. Because the isovector
$1/\Lambda_A^2$ terms contribute at the few percent level to
spin-dependent WIMP scattering, we do not attempt to model these and
neglect higher-order isoscalar current contributions.

\section{Two-body currents}

At order $Q^3$, two-body (2b) currents
enter in chiral EFT~\cite{Park}. For spin-independent WIMP scattering,
the potential impact of meson-exchange currents has been pointed out
in Ref.~\cite{Prezeau}. In chiral EFT, the long-range parts of 2b
currents are predicted. Because of their pion-exchange nature, they only
have an isovector part, ${\bf J}_{\rm 2b} = \sum^A_{i<j} {\bf J}^3_{ij}$.
For the axial-vector weak neutral current, this leads to
\begin{multline}
{\bf J}^3_{12} = - \frac{g_A}{2 F^2_{\pi}} \frac{1}{m^2_{\pi}+k^2}
\biggl[ \Bigl(c_4+\frac{1}{4 m} \Bigr) \, {\bf k} \times
({\bm \sigma}_{\times} \times {\bf k}) \, \tau^3_{\times} \\
+ 2 c_3 {\bf k} \cdot ({\bm \sigma}_1 \tau_1^3 + {\bm \sigma}_2
\tau_2^3) {\bf k} - \frac{i}{2m} {\bf k} \cdot ({\bm \sigma}_1-{\bm
\sigma}_2) {\bf q} \, \tau^3_{\times} \biggr] ,
\label{2b}
\end{multline}
where $\tau_{\times}^3=(\tau_1\times\tau_2)^3$ and the same for ${\bm
\sigma}_{\times}$, ${\bf k}=\frac{1}{2}({\bf p}'_2-{\bf p}_2-{\bf
p}'_1+{\bf p}_1)$ and ${\bf q}=\frac{1}{4}({\bf p}_1+{\bf p}'_1-{\bf
p}_2-{\bf p}'_2)$. The low-energy couplings $c_3, c_4$
relate different processes and are determined in the
pion-nucleon or nucleon-nucleon systems~\cite{RMP}.

Following Ref.~\cite{Javier}, we include the normal-ordered 1b parts of
chiral 2b currents. This is expected to be a very good approximation
in medium-mass and heavy nuclei, because of phase-space
arguments~\cite{Fermi}. We sum the second nucleon over occupied states
in a spin and isospin symmetric reference state or core:
${\bf J}^{\rm eff}_{i,{\rm 2b}} = \sum_j (1-P_{ij}) {\bf J}^3_{ij}$,
where $P_{ij}$ is the exchange operator.  Taking a Fermi gas
approximation for the core we obtain the normal-ordered 1b
current. This contains two terms. First, it leads to an in-medium
renormalization of the axial coupling~\cite{Javier}:
\begin{equation}
{\bf J}^{\rm eff}_{i,{\rm 2b}} = 
- g_A {\bm \sigma}_i \, \frac{\tau_i^3}{2} \frac{\rho}{F^2_\pi} \,
I(\rho,P=0) \biggl( \frac{1}{3} \, (2c_4-c_3) + \frac{1}{6 m} \biggr) \,,
\label{GT_1beff}
\end{equation}
with total momentum ${\bf P}={\bf p}_i+{\bf p}_i'$.
The $P$ dependence is very weak ($<15\%$) from $P=0$ to the Fermi-gas
mean-value $P^2 = 6 \kf^2/5$,
and we therefore take $P=0$ in the following.
Here, $\rho = 2\kf^3/(3\pi^2)$ is the density of the reference state, $\kf$ the
corresponding Fermi momentum, and $I(\rho,P)$ is due to the summation
in the exchange term,
\begin{equation}
I(\rho,P=0) = 
1 - \frac{3 m^2_\pi}{\kf^2}
+ \frac{3 m^3_\pi}{2 \kf^3} \, \mathrm{arccot} 
\biggl[\frac{m_\pi^2-\kf^2}{2 m_\pi \kf}\biggr] \,.
\end{equation}
Such a renormalization is also expected considering chiral 3N forces
as density-dependent two-body interactions~\cite{Holt}. We take a
typical range for the densities in nuclei $\rho= 0.10...0.12
\fmiq$. This leads to $I(\rho,P=0) = 0.58...0.60$.
The resulting 2b contribution to the
axial-vector WIMP current can be included as a density-dependent
renormalization $a_1 (1 + \delta a_1)$, with
\begin{equation}
\delta a_1 \equiv - \frac{\rho}{F^2_\pi} \, 
I(\rho,P=0) \biggl( \frac{1}{3} \, (2c_4-c_3) + \frac{1}{6 m} \biggr) \,.
\end{equation}

Normal ordering leads to a second contribution which renormalizes the
pseudo-scalar coupling. At $P=0$, this takes the form for the weak 
neutral current
\begin{equation}
{\bf J}^{{\rm eff},\, P}_{i,{\rm 2b}} = - g_A \tau_i^3 \frac{\rho}{F^2_\pi} \,
c_3 \frac{({\bf p} \cdot {\bm \sigma}_{i}) \, {\bf p}}{4 m^2_{\pi}+p^2} \,.
\label{P_1beff}
\end{equation}
For neutrinoless double-beta decay, we approximated
$({\bf p} \cdot {\bm \sigma}_{i}) \, {\bf p} \to p^2/3 \, {\bm
\sigma}_{i}$, which neglects the small tensor part between
$0^+$ ground states~\cite{Javier}. For the axial-vector WIMP
current, we fully include the contribution of Eq.~(\ref{P_1beff})
via a density- and momentum-dependent modification of the
pseudo-scalar part. To this end, we define
\begin{equation}
\delta a^P_1(p^2) \equiv -2 c_3 \, \frac{\rho}{F^2_\pi} 
\frac{p^2}{4 m^2_{\pi}+p^2} \,.
\label{deltaa1P}
\end{equation}

In addition to the long-range pion-exchange currents, there are
short-range 2b currents both for the isoscalar and isovector parts,
which are included as contact terms in chiral EFT. For Gamow-Teller
transitions, which are an isospin rotation of the axial-vector WIMP
currents, the contributions from long-range 2b currents were found to
dominate over the short-range parts in medium-mass to heavy nuclei
(for typical short-range couplings $c_D$), because $c_3, c_4$ are
large in chiral EFT without explicit Deltas~\cite{Javier}. Therefore,
we neglect the short-range 2b current contributions here, consistently
with neglecting higher-order 1b isoscalar currents, as discussed
above.

\section{WIMP-nucleus scattering}

The differential cross-section for
spin-dependent WIMP elastic scattering on a nucleus in the ground
state with total angular momentum $J$ is given by~\cite{IJMPE}
\begin{equation}
\frac{d\sigma}{dp^2} = \frac{8 G_{\rm F}^2}{(2J+1) v^2} \, S_A(p) \,,
\end{equation}
where $v$ is the WIMP velocity and $S_A(p)$ is the axial-vector
structure factor, which can be decomposed as
\begin{equation}
S_A(p) = \sum_{L \, {\rm odd}} \Bigl( \bigl|\langle J || 
{\mathcal T}_L^{{\rm el} \, 5}(p) || J \rangle \bigr|^2 
+ \bigl|\langle J || {\mathcal L}_L^{5}(p) || 
J \rangle \bigr|^2 \Bigr) \,.
\end{equation}
The sum is over multipoles $L$ with reduced matrix elements of the
transverse electric ${\mathcal T}_L^{{\rm el} \, 5}$ and longitudinal
${\mathcal L}_L^{5}$ projections of the axial-vector currents.
These can be expanded in vector spherical harmonics~\cite{IJMPE,Walecka},
leading to
\begin{align}
&{\mathcal T}_L^{{\rm el} \, 5}(p) = \frac{1}{\sqrt{2L+1}}
\sum_{i=1}^A \frac{1}{2} \Bigl[ a_0 + a_1 \tau_i^3 \Bigl(1-2 \,
\frac{p^2}{\Lambda_{A}^2} + \delta a_1 \Bigr) \Bigr] \nonumber \\[1mm]
&\times \Bigl[
- \sqrt{L} \, M_{L,L+1}(p {\bf r}_i) + \sqrt{L+1} \, M_{L,L-1}(p {\bf
r}_i) \Bigr] \,, \\[1mm]
&{\mathcal L}_L^{5}(p) = \frac{1}{\sqrt{2L+1}} 
\sum_{i=1}^A \nonumber \\[1mm]
&\times \frac{1}{2} \biggl[ a_0 + a_1 \tau_i^3
\biggl( 1 + \delta a_1 - \frac{2 g_{\pi p n} F_\pi \, p^2}{2 m g_A
(m_\pi^2+p^2)} + \delta a_1^P(p^2) \biggr) \biggr] \nonumber \\[1mm]
&\times \Bigl[\sqrt{L+1} \, M_{L,L+1}(p {\bf r}_i) 
+ \sqrt{L} \, M_{L,L-1}(p {\bf r}_i) \Bigr] \,.
\end{align}
Expressions for $M_{L,L'}(p {\bf r}_i) = j_{L'}(p r_i)
[Y_{L'}(\hat{\bf r}_i) \, {\bm \sigma}_i]^L$ (with $L'$ and ${\bm
\sigma}$ coupled to $L$) are given in Ref.~\cite{Walecka}. The
pseudo-scalar currents do not contribute to the transverse electric
multipoles, and for the longitudinal part, one can replace $({\bf p}
\cdot {\bm \sigma}_i) \, {\bf p} \to p^2 {\bm \sigma}_i$. As a result, the
$p^2/\Lambda_A^2$ terms in the longitudinal response cancel [see
Eq.~(\ref{gP})] and only contribute to the transverse electric multipoles.

\section{Nuclear structure and results}

We carry out state-of-the-art
large-scale shell-model calculations for the structure of $^{129}$Xe
and $^{131}$Xe. The valence space for both protons and neutrons
comprises the $0g_{7/2}$, $1d_{5/2}$, $1d_{3/2}$, $2s_{1/2}$, and
$0h_{11/2}$ orbitals on top of a $^{100}$Sn core. In the case of
$^{129}$Xe, the number of particles in the $1d_{3/2}$, $2s_{1/2}$, and
$0h_{11/2}$ orbitals was limited to 3, in order to make the
calculations feasible (the matrix dimension for this space is $3.5
\times 10^8$). We have used the so-called gcn5082
interaction~\cite{Menendez}, which is based on a G-matrix with
empirical adjustments, mainly in the monopole part, to describe nuclei
within this region. The same interaction and valence space have been
used to study structure and double-beta
decays. Calculations have been performed with the
shell-model code ANTOINE~\cite{Antoine}.

A fully consistent treatment of the one- and two-body currents would
require to renormalize them to the valence space of the many-body
calculation, which can lead to additional contributions to the
currents. This will be pursued in future work. In this first
application, we focus on the effects of the bare chiral currents.

\begin{figure}[t]
\begin{center}
\includegraphics[width=0.48\textwidth,clip=]{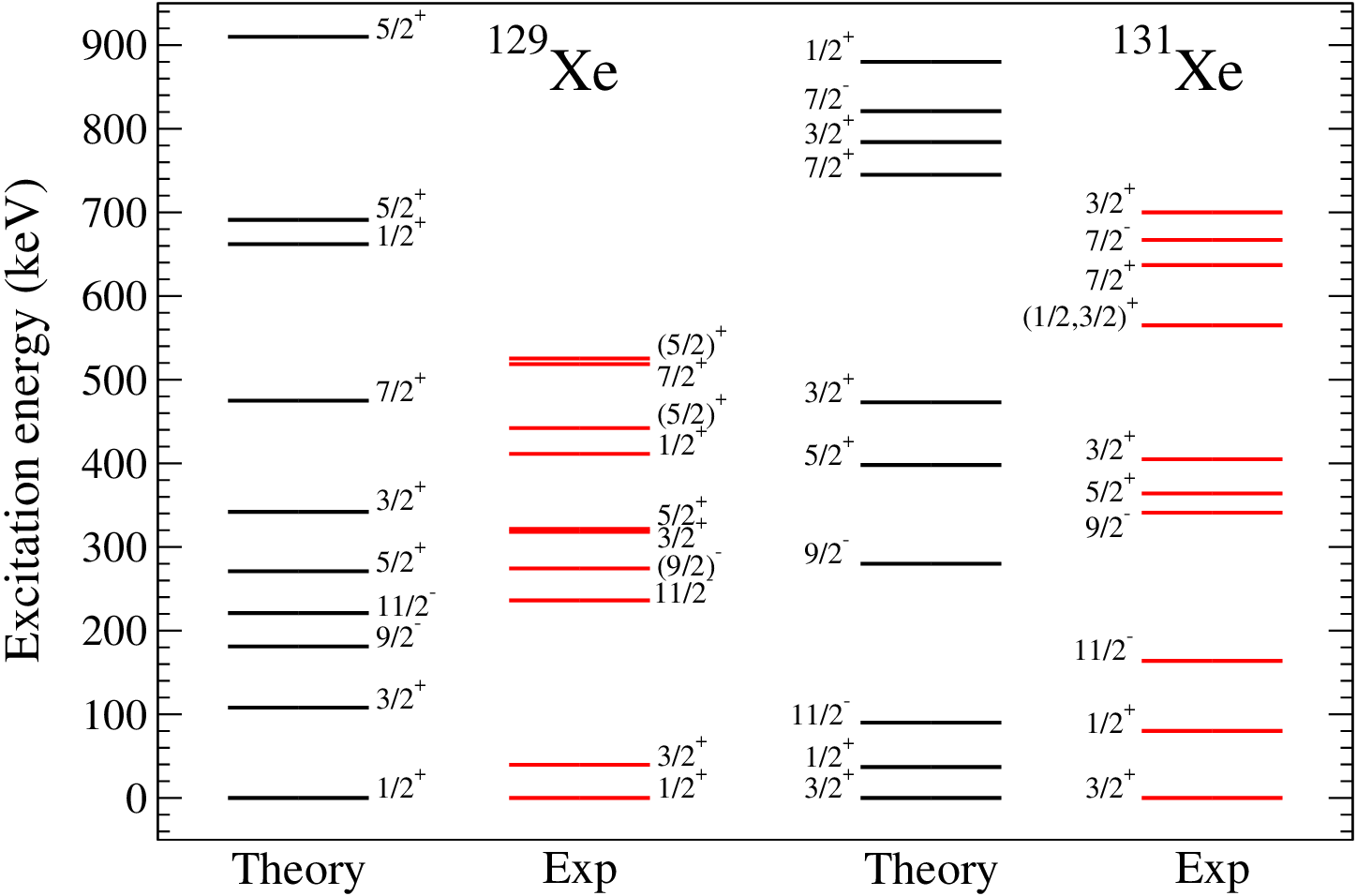}
\end{center}
\caption{(Color online) Comparison of calculated spectra of
$^{129}$Xe and $^{131}$Xe with experiment. The interaction and
valence space are the same as in structure and double-beta decay
studies in this region~\cite{Menendez}. The $^{131}$Xe 
diagonalization is in the full space, while the $^{129}$Xe
case was restricted (see text).\label{spectra}}
\end{figure}

Figure~\ref{spectra} shows the resulting spectra for $^{129}$Xe and
$^{131}$Xe, measured from the ground-state energy. In both cases, the
experimental ground state and the overall ordering of the excited
states are remarkably well described, which represents a clear
improvement with respect to previous work~\cite{Toivanen}. This
reflects the quality of the interaction and valence space used. The
resulting magnetic moments, with standard $g$ factors for this
region~\cite{Sieja}, are $\mu = -0.72 \mu_N$ and $0.86 \mu_N$ for
$^{129}$Xe and $^{131}$Xe, respectively, which agree within $\sim 20
\%$ with experiment. However, we emphasize that the magnetic moments
probe different physics, as the axial-vector WIMP couplings at $p=0$
are an isospin rotation of the Gamow-Teller operator, not a coupling
to magnetic moments. In fact, one could fine-tune the effective $g$
factors to improve the agreement with magnetic moments (which is
sometimes pursued~\cite{Toivanen}), but this leaves the predicted
axial-vector structure factor unchanged.

\begin{figure}[t]
\begin{center}
\includegraphics[width=0.48\textwidth,clip=]{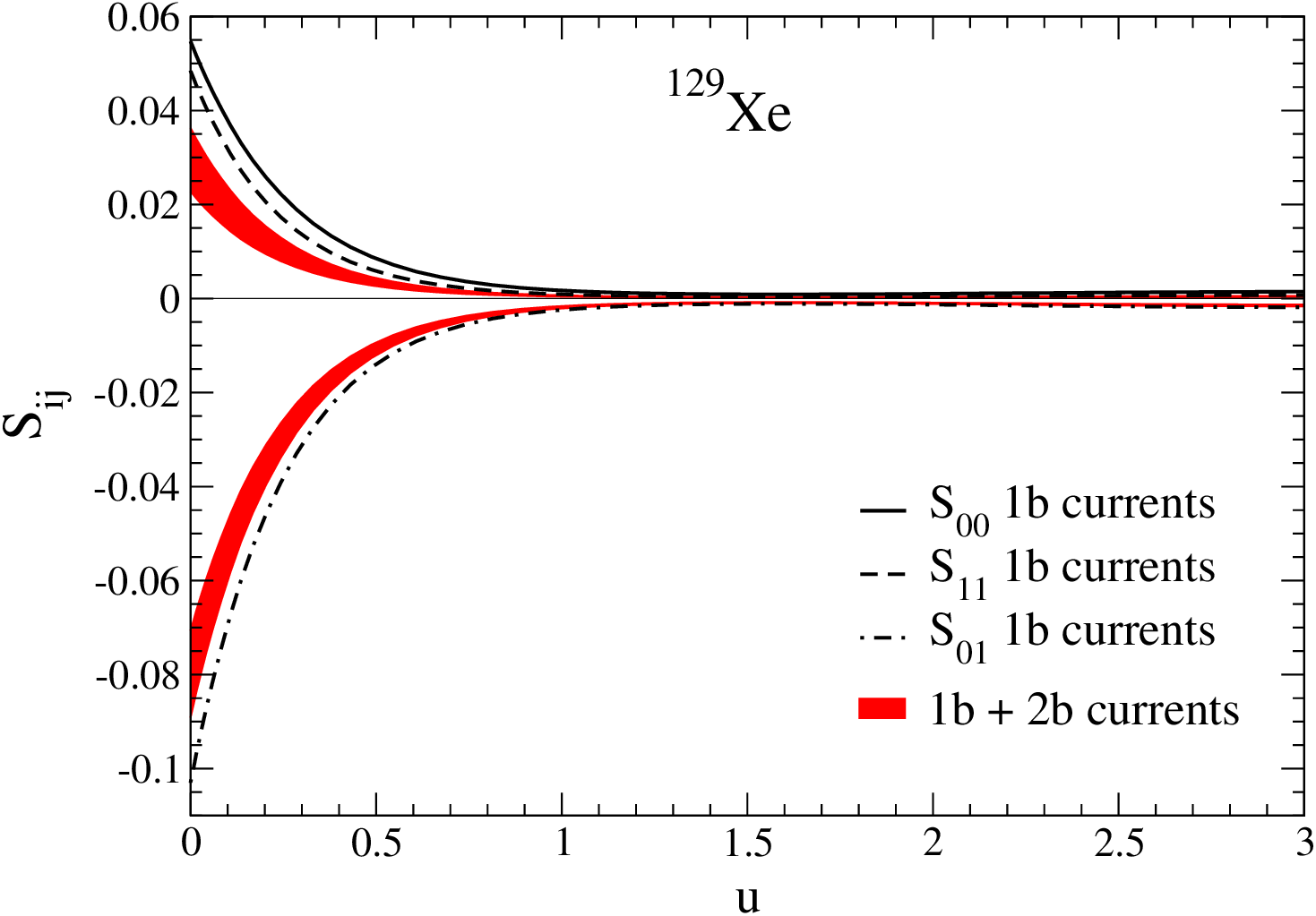}
\end{center}
\caption{(Color online) Structure factors $S_{00}$, $S_{11}$ and
$S_{01}$ for $^{129}$Xe as a function of $u=p^2b^2/2$, with
harmonic-oscillator length $b= 2.2853\, {\rm fm}$. Results are shown at
the one-body (1b) current level, and for the isovector
and isoscalar-isovector factors $S_{11}$ and
$S_{01}$ also including two-body (2b) currents.
The estimated theoretical uncertainty is given by the
red band.\label{Xe129_Sij}}
\end{figure}

In the limit of low momentum-transfer, $p=0$, the axial-vector
structure factor reduces to the expectation values of the total
proton and neutron spin operators, ${\bf S}_p = \sum_{i=1}^Z {\bm
\sigma}_i/2$ and ${\bf S}_n = \sum_{i=1}^N {\bm \sigma}_i/2$,
\begin{equation}
S_A(0) = \frac{1}{4 \pi} \bigl| (a_0+a_1') \langle J || {\bf S}_p ||
J \rangle + (a_0-a_1') \langle J || {\bf S}_n || J \rangle \bigl|^2 \,,
\end{equation}
where $a_1'=a_1(1+\delta a_1)$ includes the effects from chiral
2b currents. For the expectation values $\langle {\bf S}_{n,p}
\rangle = \langle J M=J | {\bf S}^3_{n,p} | J M=J \rangle$, we obtain
\begin{equation}
\begin{array}{c|c|c}
& \: ^{129}{\rm Xe} \: & \: ^{131}{\rm Xe} \: \\ \hline
\: \langle {\bf S}_n \rangle \: & \: 0.329 \: & \: -0.272 \: \\ \hline
\: \langle {\bf S}_p \rangle \: & \: 0.010 \: & \: -0.009 \: 
\end{array}
\end{equation}
As expected for odd-mass nuclei with even number of protons, $\langle
{\bf S}_n \rangle \gg \langle {\bf S}_p \rangle$. The predicted spin
expectation values are qualitatively similar to those of Ressell and
Dean~\cite{Dean}, who considered two different (Bonn A/Nijmegen II)
interactions, although their $\langle {\bf S}_p \rangle=0.013/0.028$
for $^{129}$Xe is larger and their $\langle {\bf S}_n \rangle$ values
for $^{131}$Xe are $20\%$ smaller. The results of Toivanen {\it et
  al.}~\cite{Toivanen} differ substantially, as expected based on
their poor reproduction of the spectra. They give $15\%/55\%$ smaller
$\langle {\bf S}_n \rangle$ values for $^{129}$Xe/$^{131}$Xe and both
$\langle {\bf S}_p \rangle$ values an order of magnitude smaller (for
$^{129}$Xe also of opposite sign). We attribute these differences to
the sizeable truncations of the valence spaces in
Refs.~\cite{Dean,Toivanen} and also because the nuclear interactions
are not as well tested compared to this work.

\begin{figure}[t]
\begin{center}
\includegraphics[width=0.48\textwidth,clip=]{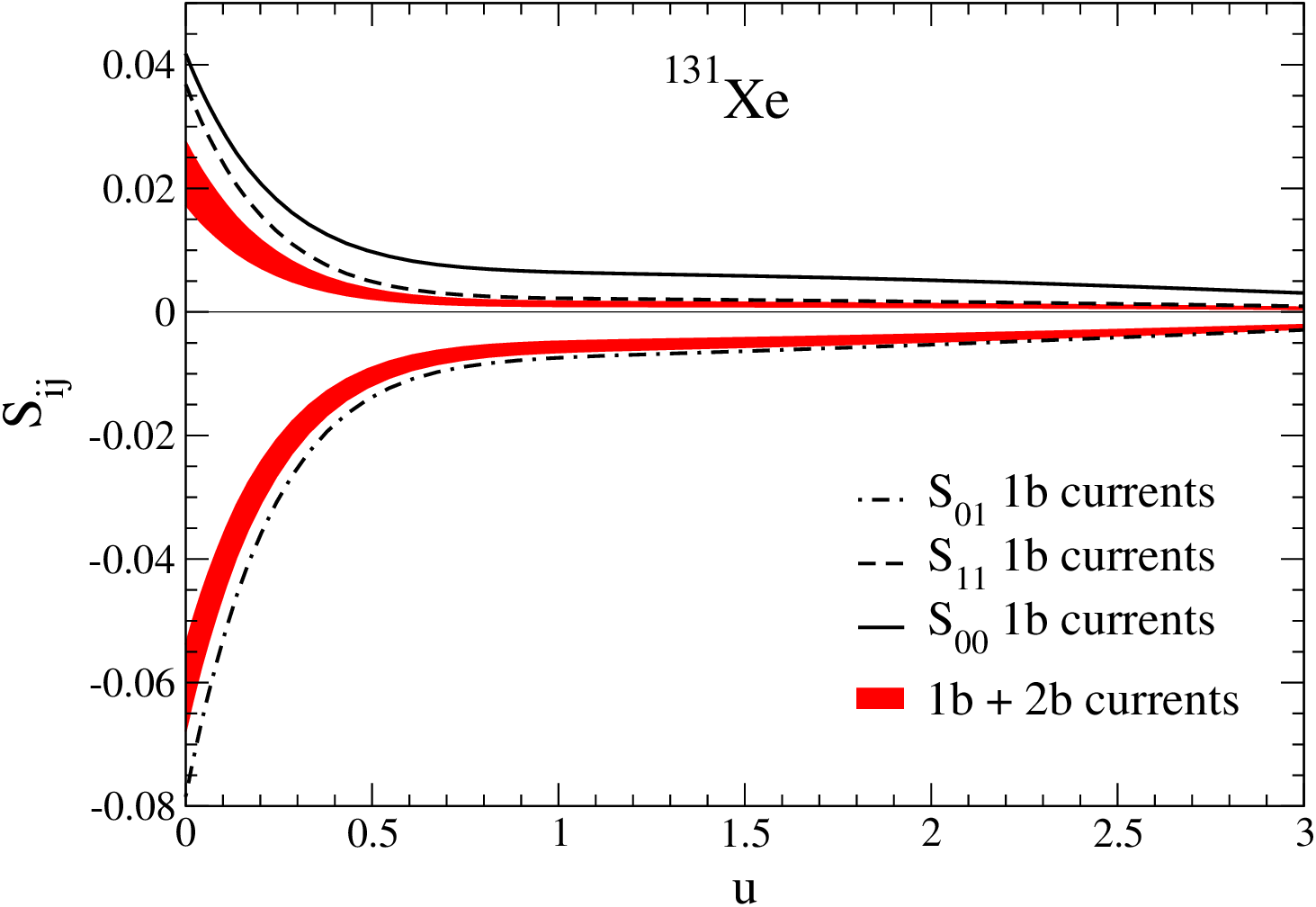}
\end{center}
\caption{(Color online) Structure factors $S_{00}$, $S_{11}$ and
$S_{01}$ for $^{131}$Xe as a function of $u=p^2b^2/2$, with
harmonic-oscillator length $b= 2.2905\, {\rm fm}$. The labels are
as in Fig.~\ref{Xe129_Sij}.\label{Xe131_Sij}}
\end{figure}

At finite $p$, one introduces isoscalar/isovector (0/1) structure
factors $S_{00}(p)$, $S_{01}(p)$ and $S_{11}(p)$ through
\begin{equation}
S_A(p) = a_0^2 \, S_{00}(p) + a_0 a_1 S_{01}(p) + a_1^2 \,
S_{11}(p) \,.
\end{equation}
Figures~\ref{Xe129_Sij} and~\ref{Xe131_Sij} show the predicted
structure factors for $^{129}$Xe and $^{131}$Xe as a function of
$u=p^2b^2/2$, with harmonic-oscillator length $b=(\hbar/m
\omega)^{1/2}$ and $\hbar\omega = (45 A^{-1/3} - 25 A^{-2/3})
\mev$. Fits to $S_{00}$, $S_{11}$ and $S_{01}$ are provided in Table~\ref{fits}.

At the 1b level at order $Q^2$, the results shown correspond to the
same current operator as in Refs.~\cite{Dean,Toivanen}, but based on
chiral EFT couplings and with the $1/\Lambda_A^2$ term in $g_A$ and
$g_P$ included. Since $\langle {\bf S}_n \rangle$ sets the scale for
the $p=0$ response, one has $S_{00} \approx S_{11} \approx -1/2 \,
S_{01}$. Thus, our response is about $35 \%$ larger for $^{131}$Xe
compared to Ressell and Dean~\cite{Dean}, and about $30 \%$/4 times
larger for $^{129}$Xe/$^{131}$Xe than Toivanen {\it et
al.}~\cite{Toivanen}.  For $p \gtrsim m_\pi$, $u \gtrsim 1.3$, the
structure factors nearly vanish for $^{129}$Xe, similar to
Refs.~\cite{Dean,Toivanen}, as expected for a $1/2^+$ ground state,
where only $L=1$ multipoles contribute. For $^{131}$Xe, the predicted
structure factors remain nonvanishing but small, similar to Ressell
and Dean~\cite{Dean}. Both differ significantly from Toivanen {\it et
al.}~\cite{Toivanen} suggesting very different higher $L=3$
multipole contributions.

When chiral 2b currents are included at order $Q^3$, we provide
theoretical error bands in Figs.~\ref{Xe129_Sij} and~\ref{Xe131_Sij}
due to the nuclear uncertainties in WIMP currents in nuclei. This
takes into account the uncertainties in the low-energy couplings $c_3,
c_4$ (with the conservative range used in Ref.~\cite{Javier}) as well
as the range in nuclear densities $\rho= 0.10...0.12 \fmiq$. Chiral 2b
currents provide important contributions to the structure factors,
especially for $p \lesssim 100 \mev$, $u \lesssim 0.7$, where their
effects are to significantly reduce, by about $25-55 \%$, the isovector
$S_{11}$ factor, similar to the Gamow-Teller quenching~\cite{Javier}
and by about half of that the isoscalar-isovector $S_{01}$ factor,
compared to the results at the 1b level at order $Q^2$.

For $p \gtrsim m_{\pi}$, the effects of 2b currents are expected to be
smaller due to the momentum-dependent modification of the
pseudo-scalar part of the current, $\delta a^P_1(p^2)$, which has
opposite sign compared to the leading 1b contribution (see Eq.~(10)
with $c_3<0$~\cite{Javier}). This term, however, only affects the
longitudinal multipoles, leading to a nucleus-dependent reduction. For
$^{129}$Xe, most of the response comes from the transverse electric
multipoles, so that the relative contribution from chiral 2b currents
is approximately $p$ independent.  For $^{131}$Xe, at $p \sim 250
\mev$ the relative effect of chiral 2b currents is $\sim 10-15\%$
smaller than at $p \sim 100 \mev$, in agreement with the
momentum-transfer dependence of chiral 2b currents found in
Gamow-Teller transitions~\cite{Javier}.

As discussed,
the 2b current contributions to the isoscalar $S_{00}$ factor are
small and are not considered here. The chiral 2b current contributions
should be included in all calculations of spin-dependent elastic and
inelastic WIMP scattering off nuclei. The effects of 2b currents are
expected to be smaller for the vector structure factor and thus for
the spin-independent elastic WIMP response.

\section{Summary}

This presents the first calculation of spin-dependent
WIMP currents in nuclei based on chiral EFT, including the leading
long-range 2b currents. They predict a $25-55 \%$ reduction of the
isovector part of the one-body axial-vector WIMP currents, where the
range provides an estimate of the theoretical uncertainties of WIMP
currents in nuclei. This should be included in limits on the WIMP
couplings, where the spin-dependent analysis provides complementary
constraints~\cite{exp}. As an application, we have calculated the
structure factors for spin-dependent WIMP scattering off
$^{129,131}$Xe nuclei, using the largest valence spaces accessible with
nuclear interactions that have been tested in nuclear structure and
double-beta decay studies in this region. Future work includes
developing consistent interactions based on chiral EFT, where the
present frontier is in the calcium region~\cite{Gallant}, and
investigating other nuclei and responses.

\section*{Acknowledgments}

We thank L.\ Baudis, T.\ Marrodan and U.-G.\ Mei{\ss}ner for helpful
discussions. This work was supported in part by the DFG through grant
SFB 634, the Helmholtz Alliance HA216/EMMI, and a BMBF ARCHES Award.

\newpage

\onecolumngrid

\begin{table}[t]
\caption{Fits to the structure factors $S_{00}$, $S_{11}$ and $S_{01}$
for spin-dependent WIMP elastic scattering off $^{129}$Xe and
$^{131}$Xe nuclei, including 1b and 2b currents as in
Figs.~\ref{Xe129_Sij} and~\ref{Xe131_Sij}. For the 1b+2b current results,
both the central value of the theoretical error band was used for the
fits (first rows) and the limits of the band (second rows).
The fitting function of the dimensionless variable $u = p^2
b^2/2$ is $S_{ij}(u) =
e^{-u} \sum_{n=0}^9 c_{ij,n} u^n$.
The rows give the coefficients $c_{ij,n}$ of the
$u^n$ terms in the polynomial.\label{fits}}
\begin{center}
\begin{tabular*}{0.765\textwidth}{c||c|c|c|c|c}
\hline
\multicolumn{6}{c}{$^{129}$Xe} \\
\multicolumn{6}{c}{$u=p^2b^2/2 \,, \: b=2.2853 \, {\rm fm}$} \\
\hline
$e^{-u}\times$ & $S_{00}$ & $S_{11}$ (1b) & $S_{11}$ (1b+2b) & 
$S_{01}$ (1b) & $S_{01}$ (1b+2b) \\
\hline
$1$ & $0.054731$ & $0.048192$ & $0.02933$ & $-0.102732$ & $-0.0796645$ \\
$u$ & $-0.146897$ & $-0.148361$ & $-0.0905396$ & $0.297105$ & $0.231997$ \\
$u^2$ & $0.182479$ & $0.202347$ & $0.122783$ & $-0.387513$ & $-0.304198$ \\
$u^3$ & $-0.128112$ & $-0.151853$ & $-0.0912046$ & $0.281816$ & $0.222024$ \\
$u^4$ & $0.0539978$ & $0.0674284$ & $0.0401076$ & $-0.122388$ & $-0.096693$ \\
$u^5$ & $-0.0133335$ & $-0.0179342$ & $-0.010598$ & $0.0317668$ & $0.0251835$ \\
$u^6$ & $0.00190579$ & $0.00286368$ & $0.00168737$ & $-0.00492337$ & 
$-0.00392356$ \\
$u^7$ & $-1.48373\times10^{-4}$ & $-2.65795\times10^{-4}$ &
$-1.56768\times10^{-4}$ & $4.39836\times10^{-4}$ & $3.53343\times10^{-4}$ \\
$u^8$ & $5.11732\times10^{-6}$ & $1.29656\times10^{-5}$ &
$7.69202\times10^{-6}$ & $-2.02852\times10^{-5}$ & $-1.65058\times10^{-5}$ \\
$u^9$ & $-2.06597\times10^{-8}$ & $-2.47418\times10^{-7}$ &
$-1.48874\times10^{-7}$ & $3.46755\times10^{-7}$ & $2.88576\times10^{-7}$ \\
\hline
$e^{-u}\times$ & & \multicolumn{2}{c|}{$S_{11}$ (1b+2b band)} & 
\multicolumn{2}{c}{$S_{01}$ (1b+2b band)} \\
\hline
$1$ && $0.0360513$ & $0.0226064$ & $-0.0888962$ & $-0.070431$ \\
$u$ && $-0.110705$ & $-0.0703558$ & $0.257562$ & $0.20644$ \\
$u^2$ && $0.150026$ & $0.0954932$ & $-0.336681$ & $-0.271749$ \\
$u^3$ && $-0.111714$ & $-0.0706386$ & $0.245328$ & $0.198772$ \\
$u^4$ && $0.0493115$ & $0.0308683$ & $-0.106746$ & $-0.0866783$ \\
$u^5$ && $-0.0130745$ & $-0.00810917$ & $0.0277603$ & $0.0226217$ \\
$u^6$ && $0.00208705$ & $0.00128522$ & $-0.00431334$ & $-0.00353705$ \\
$u^7$ && $-1.94174 \times 10^{-4}$ & $-1.19086 \times 10^{-4}$
& $ 3.86604 \times 10^{-4}$ & $3.20472 \times 10^{-4}$ \\
$u^8$ && $9.52295 \times 10^{-6}$ & $5.84562 \times 10^{-6}$ &
$-1.79013 \times 10^{-5}$ & $-1.51335 \times 10^{-5}$ \\
$u^9$ && $-1.83523 \times 10^{-7}$ & $-1.13885 \times 10^{-7}$ &
$3.06893 \times 10^{-7}$ & $2.70785 \times 10^{-7}$ \\
\hline
\hline
\multicolumn{6}{c}{$^{131}$Xe} \\
\multicolumn{6}{c}{$u=p^2b^2/2 \,, \: b=2.2905 \, {\rm fm}$} \\
\hline 
$e^{-u}\times$ & $S_{00}$ & $S_{11}$ (1b) & $S_{11}$ (1b+2b) &
$S_{01}$ (1b) & $S_{01}$ (1b+2b) \\
\hline
$1$ & $0.0417889$ & $0.0368132$ & $0.022446$ & $-0.0784478$ & $-0.0608808$ \\
$u$ & $-0.111171$ & $-0.118361$ & $-0.0733931$ & $0.230484$ & $0.181473$ \\
$u^2$ & $0.171966$ & $0.176773$ & $0.110509$ & $-0.343106$ & $-0.272533$ \\
$u^3$ & $-0.133219$ & $-0.137987$ & $-0.0868752$ & $0.263525$ & $0.211776$ \\
$u^4$ & $0.0633805$ & $0.063821$ & $0.0405399$ & $-0.120946$ & $-0.0985956$ \\
$u^5$ & $-0.0178388$ & $-0.0176743$ & $-0.0113544$ & $0.0331754$ & $0.027438$ \\
$u^6$ & $0.00282476$ & $0.00287653$ & $0.00187572$ & $-0.00528724$ & 
$-0.0044424$ \\
$u^7$ & $-2.31681\times10^{-4}$ & $-2.63605\times10^{-4}$ &
$-1.75285\times10^{-4}$ & $4.6475\times10^{-4}$ & $3.97619\times10^{-4}$ \\
$u^8$ & $7.78223\times10^{-6}$ & $1.23239\times10^{-5}$ &
$8.40043\times10^{-6}$ & $-2.00407\times10^{-5}$ & $-1.74758\times10^{-5}$ \\
$u^9$ & $-4.49287\times10^{-10}$ & $-2.17839\times10^{-7}$ &
$-1.53632\times10^{-7}$ & $2.90375\times10^{-7}$ & $2.55979\times10^{-7}$ \\
\hline
$e^{-u}\times$ & & \multicolumn{2}{c|}{$S_{11}$ (1b+2b band)} & 
\multicolumn{2}{c}{$S_{01}$ (1b+2b band)} \\
\hline
$1$ & & $0.0275713$ & $ 0.0173191$ & $-0.067911$ & $-0.0538485$ \\
$u$ & & $ -0.0891521$ & $-0.0576174$ & $ 0.20071 $ & $ 0.162246$ \\
$u^2$ & & $0.134094 $ & $ 0.0868883$ & $-0.301325$ & $-0.24379$ \\
$u^3$ & & $ -0.105277 $ & $-0.0684403$ & $ 0.233566$ & $ 0.190041$ \\
$u^4$ & & $0.0490864$ & $ 0.0319771$ & $-0.108494$ & $-0.0887141$ \\
$u^5$ & & $ -0.0137279$ & $-0.00897632 $ & $ 0.0301064 $ & $ 0.0247694$ \\
$u^6$ & & $0.00226103 $ & $ 0.00148964 $ & $-0.00485087$ & $-0.0040324$ \\
$u^7$ & & $-2.10252 \times 10^{-4}$ & $-1.40246 \times 10^{-4}$ & 
$4.30628 \times 10^{-4}$ & $3.64368 \times 10^{-4}$ \\
$u^8$ & & $1.0004 \times 10^{-5}$ & $6.79325 \times 10^{-6}$ &
$-1.86527 \times 10^{-5}$ & $-1.62836 \times 10^{-5}$ \\
$u^9$ & & $-1.80793 \times 10^{-7}$ & $-1.26396 \times 10^{-7}$ &
$2.63842 \times 10^{-7}$ & $2.48126 \times 10^{-7}$ \\
\hline
\end{tabular*}
\end{center}
\end{table}


\begin{thebibliography}{99}
\bibitem{theo} R.\ J.\ Gaitskell,
Annu.\ Rev.\ Nucl.\ Part.\ Sci.\ {\bf 54}, 315 (2004);
J.\ L.\ Feng,
Annu.\ Rev.\ Astron.\ Astrophys.\ {\bf 48}, 495 (2010).

\bibitem{new} A.\ L.\ Fitzpatrick, W.\ Haxton, E.\ Katz, N.\ Lubbers
and Y.\ Xu, arXiv:1203.3542;
V.\ Cirigliano, M.\ L.\ Graesser and G.\ Ovanesyan,
arXiv:1205.2695.

\bibitem{exp} H.\ S.\ Lee {\it et al.} (KIMS Collaboration),
Phys.\ Rev.\ Lett.\ {\bf 99}, 091301 (2007);
J.\ Angle {\it et al.} (XENON10 Collaboration),
Phys.\ Rev.\ Lett.\ {\bf 101}, 091301 (2008);
V.\ N.\ Lebedenko {\it et al.} (ZEPLIN-III Collaboration),
Phys.\ Rev.\ Lett.\ {\bf 103}, 151302 (2009);
E.\ Behnke {\it et al.} (COUPP Collaboration),
Phys.\ Rev.\ Lett.\ {\bf 106}, 021303 (2011);
S.\ Archambault {\it et al.} (PICASSO Collaboration),
Phys.\ Lett.\ B {\bf 711}, 153 (2012);
M.\ Felizardo {\it et al.} (SIMPLE Collaboration),
Phys.\ Rev.\ Lett.\ {\bf 108}, 201302 (2012).

\bibitem{SNO} M.\ Butler, J.-W.\ Chen and X.\ Kong, Phys.\ Rev.\ C
{\bf 63}, 035501 (2001); S.\ Nakamura, T.\ Sato, V.\ P.\ Gudkov and 
K.\ Kubodera, Phys.\ Rev.\ C {\bf 63}, 034617 (2001).

\bibitem{Javier} J.\ Men\'{e}ndez, D.\ Gazit and A.\ Schwenk,
Phys.\ Rev.\ Lett.\ {\bf 107}, 062501 (2011).

\bibitem{XENON} E.\ Aprile {\it et al.} (XENON100 Collaboration),
Phys.\ Rev.\ Lett.\ {\bf 107}, 131302 (2011) and arXiv:1207.5988.

\bibitem{Menendez} E.\ Caurier, J.\ Men\'{e}ndez, F.\ Nowacki and
A.\ Poves, Phys.\ Rev.\ Lett.\ {\bf 100}, 052503 (2008);
J.\ Men\'{e}ndez, A.\ Poves, E.\ Caurier and F.\ Nowacki, 
Nucl.\ Phys.\ A {\bf 818}, 139 (2009).

\bibitem{IJMPE} J.\ Engel, S.\ Pittel and P.\ Vogel, Int.\ J.\
Mod. Phys. E {\bf 1}, 1 (1992).

\bibitem{Kam} G.\ Jungman, M.\ Kamionkowski and K. Griest,
Phys.\ Rep.\ {\bf 267}, 195 (1996).

\bibitem{Bernard} V.\ Bernard, L.\ Elouadrhiri and U.-G.\ 
Mei{\ss}ner, J.\ Phys.\ G {\bf 28}, R1 (2002).

\bibitem{isoscalar} C.\ E.\ Carlson and J.\ L.\ Poor, Phys.\ Rev.\
D {\bf 36}, 2169 (1987).

\bibitem{Park} T.\ S.\ Park, L.\ E.\ Marcucci, R.\ Schiavilla,
M.\ Viviani, A.\ Kievsky, S.\ Rosati, K.\ Kubodera, D.-P.\ Min
and M.\ Rho, Phys.\ Rev.\ C {\bf 67}, 055206 (2003).

\bibitem{Prezeau} G.\ Pr\'{e}zeau, A.\ Kurylov, M.\ Kamionkowski
and P.\ Vogel, Phys.\ Rev.\ Lett.\ {\bf 91}, 231301 (2003).

\bibitem{RMP} E.\ Epelbaum, H.-W.\ Hammer and U.-G.\ Mei{\ss}ner,
Rev.\ Mod.\ Phys.\ {\bf 81}, 1773 (2009).

\bibitem{Fermi} B.\ Friman and A.\ Schwenk, in {\it From Nuclei to 
Stars}, Festschrift in Honor of Gerald E.\ Brown, Ed.\ S.\ Lee 
(World Scientific, 2011), arXiv:1101.4858.

\bibitem{Holt} J.\ W.\ Holt, N.\ Kaiser and W.\ Weise, Phys.\
Rev.\ C {\bf 79}, 054331 (2009); {\it ibid.} {\bf 81}, 024002 (2010).


\bibitem{Walecka} J.\ D.\ Walecka, {\it Theoretical Nuclear and 
Subnuclear Physics} (Oxford University Press, 1995), see p.~119.

\bibitem{Antoine} E.\ Caurier, G.\ Mart\'{i}nez-Pinedo, F.\ Nowacki,
A.\ Poves and A.\ P.\ Zuker, Rev.\ Mod.\ Phys.\ {\bf 77}, 427 (2005).

\bibitem{Toivanen} P.\ Toivanen, M.\ Kortelainen, J.\ Suhonen
and J.\ Toivanen, Phys.\ Rev.\ C {\bf 79}, 044302 (2009).

\bibitem{Sieja} K.\ Sieja, G.\ Mart\'{i}nez-Pinedo, L.\ Coquard
and N.\ Pietralla, Phys.\ Rev.\ C\ {\bf 80}, 054311 (2009).

\bibitem{Dean} M.\ T.\ Ressell and D.\ J.\ Dean, Phys.\ Rev.\ C
{\bf 56}, 535 (1997).

\bibitem{Gallant} A.\ T.\ Gallant {\it et al.},
Phys.\ Rev.\ Lett.\ {\bf 109}, 032506 (2012) and references therein.
\end{thebibliography}
\end{document}